\documentclass[journal]{IEEEtran}

\usepackage{graphicx}
\usepackage{amsmath,amssymb}
\usepackage{booktabs}
\usepackage{multirow}
\usepackage{array}
\usepackage{url}
\usepackage{cite}
\usepackage{subcaption}
\usepackage{xcolor}
\usepackage{siunitx}
\usepackage{cuted}
\begin{document}

\title{Passive Smartphone-Microphone Phonocardiography for Joint Maternal--Fetal Cardiac Well-Being Monitoring: A Longitudinal Self-Recorded Case Study in Late Pregnancy}

\author{
    \IEEEauthorblockN{Muhammad Mahboob Ur Rahman}\\
    \IEEEauthorblockA{Electrical engineering department, Information Technology University (ITU), Lahore, Pakistan.}
    \thanks{This is a preliminary, exploratory analysis intended to motivate a properly instrumented, reference-validated follow-up study.}
}

%\markboth{DRAFT for submission -- IEEE Journal of Biomedical and Health Informatics (target venue)}{}

\maketitle

\begin{abstract}
Maternal and perinatal deaths are concentrated in low- and middle-income countries, where routine fetal monitoring is largely inaccessible. With the prevalence of smartphone as personal assistive gadget, we ask whether a phone's built-in microphone, with no extra hardware or emitted signal, can serve as a software-only maternal-fetal cardiac screening aid when moved from the chest to the lower abdomen? We report a longitudinal, self-recorded case study of a single subject (Z) over the last five weeks of a singleton pregnancy, later delivered at term without complications: 28 short, uncontrolled recordings (rest, morning, post-exercise). Lacking maternal/fetal ground truth, we exploit the prior that maternal and fetal heart rates occupy largely disjoint frequency bands, blindly decomposing a Shannon-energy envelope into two candidate-rate bands, and additionally probe a slower breathing-rate band as a rest/exercise discriminator. A fast, regular, fetal-plausible rhythm was recovered from most recordings and remained essentially unchanged across maternal activity states, consistent with normal fetal cardiac autonomy; a slower, noisier maternal-plausible rhythm was recovered less reliably. Beat-to-beat variability proved largely noise-dominated at our short window lengths, and the breathing-rate probe did not significantly separate rest from exercise; both are reported as negative findings. We position the work within AI for global health: a low-cost screening aid for remote and underserved communities, pending reference-validated follow-up, not a replacement for clinical monitoring.
\end{abstract}


\section{Introduction}
\label{sec:intro}

\subsection{A Global Health Motivation}
\label{sec:motivation}
Maternal and perinatal mortality remain among the starkest health inequities in the world. In 2023, low-income countries recorded 346 maternal deaths per 100{,}000 live births against just 10 in high-income countries, and 94\% of all maternal deaths occurred in low- and middle-income countries (LMICs), with Sub-Saharan Africa and South Asia alone accounting for 87\% of the global total \cite{un_sdg3_2026}. A large share of this burden is attributable not to untreatable disease but to limited access: skilled attendance at birth, routine antenatal screening, and continuous fetal heart-rate monitoring (CTG/Doppler) remain concentrated in well-resourced clinical settings and are frequently unavailable to rural, remote, or low-income populations \cite{un_sdg3_2026}. At the same time, mobile technology has reached a tipping point: smartphone ownership is now a global majority, and mobile phone ownership exceeds 84\% of adults even in developing economies \cite{gsma_findex_2025,worldbank_findex_2025} -- including many of the same communities underserved by routine prenatal monitoring. This juxtaposition -- a largely preventable, geographically concentrated health burden, alongside a nearly-universal computing and sensing device already in people's pockets -- is precisely the kind of problem the \emph{AI for Good} and digital-health-for-equity agendas (aligned with UN Sustainable Development Goal 3 and WHO priorities on reducing preventable maternal and newborn deaths) exist to address. This paper is motivated by a deliberately narrow, concrete instance of that larger agenda: can a smartphone's already-present microphone, with no additional hardware, specialized probe, or connectivity requirement, be turned into a software-only maternal-fetal cardiac screening aid -- something closer to a free accessory than a \$2{,}000 Doppler unit? We do not claim to answer this at scale; we report a rigorous single-subject feasibility case study intended to responsibly scope what such a tool can and cannot currently do, as a step toward that goal rather than a finished solution.

\subsection{Technical Background}
The stethoscope has been the bedside instrument of cardiac assessment for two centuries, and in the last decade a substantial body of work has shown that an ordinary smartphone microphone -- with no specialized transducer -- can substitute for it with surprising fidelity when placed directly against the chest wall \cite{fonocheck2025,digitalstethoscope2017,textilecoupling2026}. Commercial and research applications built around this observation (FonoCheck, iStethoscope, Echoes, and various low-cost digital-stethoscope prototypes) have demonstrated that most of the diagnostically relevant energy in the first and second heart sounds (S1, S2) lies below approximately 250~Hz \cite{textilecoupling2026}, comfortably within the passband of a phone's built-in microphone and analogue front end.

Pregnancy poses a related but distinct sensing problem: two hearts, at very different rates and with very different acoustic transmission paths, must be monitored simultaneously from a single external vantage point on the maternal abdomen. The clinical gold standard, cardiotocography (CTG), uses a dedicated Doppler ultrasound transducer for the fetal heart rate (FHR) and a separate tocodynamometer for uterine activity \cite{comparative_fpcg_ctg_2017}; it is effective but confined to clinical settings, sensitive to maternal movement, and -- because it relies on continuous ultrasound insonation -- not considered appropriate for unsupervised, all-day, at-home use. This gap has motivated a wave of consumer-facing alternatives. Very recently, researchers proposed \emph{DopFone}, a smartphone app that turns the phone's own speaker and microphone into an active sonar-like Doppler device, emitting an ultrasonic tone and using a learned model to infer FHR from the reflected echo, reporting agreement within a few beats-per-minute of clinical Doppler on a small cohort \cite{dopfone2026}. Commercial devices such as HeraBEAT similarly pair a dedicated Doppler probe with a smartphone interface for combined maternal and fetal heart-rate display \cite{herabeat2020}. Both approaches are \emph{active} sensing systems: they emit a designed acoustic or ultrasonic probe signal and interpret its reflection or Doppler shift.

In this paper we instead ask a narrower, purely \emph{passive} question, closer in spirit to classical fetal phonocardiography (fPCG) than to Doppler sonography: can the microphone alone, listening to naturally occurring heart sounds transmitted through maternal tissue, recover usable information about \emph{both} maternal and fetal cardiac rhythm, using nothing but a stock voice-recorder app and no purpose-built sensor, adhesive patch, gel, or emitted signal? Multi-channel, purpose-built fPCG systems using arrays of piezoelectric contact sensors and blind source separation (BSS) have shown that this is possible with dedicated hardware and 3--4 simultaneous channels \cite{comparative_fpcg_ctg_2017,fourchannel_validation_2018}; whether a \emph{single}, uncalibrated, consumer microphone -- pressed against the lower abdomen for a few tens of seconds by a non-expert user, in an ordinary home environment -- retains enough signal to say anything useful about either heart, let alone both, is, to our knowledge, not previously reported.

We answer this question empirically using a unique longitudinal dataset: 28 short smartphone-microphone recordings self-collected by subject Z over the final five weeks of an otherwise uncomplicated singleton pregnancy, under three loosely-controlled activity conditions (quiet rest, first thing in the morning, and immediately following light exercise). Because no simultaneous reference (CTG, ECG, or Doppler) was available, we cannot perform supervised source separation; instead we develop a signal-processing pipeline that (a) automatically locates the short, high-quality windows embedded within otherwise silent or noise-dominated recordings, and (b) exploits normal physiological heart-rate ranges as a \emph{prior} to blindly decompose the recovered signal into a slow, maternal-plausible band and a fast, fetal-plausible band. We report the resulting maternal- and fetal-rate estimates across all 28 recordings, characterize their stability across activity conditions and across the five-week span, and discuss, with explicit epistemic caution, what can and cannot be concluded from single-channel, unlabelled, ecologically-collected audio of this kind.

\subsection{Contributions}
\begin{enumerate}
\item We present, to our knowledge, the first reported attempt at passive, single-channel, consumer-smartphone-microphone acquisition of \emph{joint} maternal and fetal heart-sound information from unsupervised, self-collected, real-world recordings, rather than laboratory or clinically supervised sessions.
\item We describe a fully automated processing pipeline -- robust envelope extraction, active-segment detection, and physiologically-informed dual-band harmonic disambiguation -- designed specifically for the very low and highly variable signal-to-noise ratio (SNR) regime of unassisted self-recording.
\item We report a 28-recording, 18-day longitudinal case study spanning the last five weeks of pregnancy, showing that a fast, regular, fetal-plausible rhythm ($160.3\pm11.5$~bpm) is recoverable in 25/28 recordings and is stable across maternal activity states, while a slower maternal-plausible rhythm ($82.5\pm17.9$~bpm) is recoverable less reliably (14/28) and is confounded by exercise-induced band overlap. We further extract beat-to-beat variability (SDNN, RMSSD, and a CTG-style bandwidth metric) from both bands and show, via an explicit small-sample-size diagnostic, that most individual-window variability estimates are noise-dominated -- an honest negative finding that scopes what this acquisition modality can and cannot currently support.
\item We further probe whether a third, even-lower disjoint frequency band -- the maternal breathing rate -- can be extracted from the same recordings to distinguish rest from exercise sessions, and report, via an explicit statistical test, that it does not: an honest negative result for a plausible, low-cost extension of the method.
\item We situate the approach relative to prior smartphone PCG, multi-channel fPCG source separation, and contemporaneous active-Doppler smartphone apps, and we are explicit about the respects in which single-channel passive PCG under-determines the maternal/fetal attribution problem, to responsibly scope claims for follow-up, reference-validated work.
\item We position this work within the \emph{AI for Good} / digital-health-equity agenda: given the stark, well-documented disparity in access to routine fetal and maternal monitoring between high- and low-resource settings, and the near-universal reach of smartphone ownership even in developing economies, we argue that a validated version of this approach could plausibly serve as a zero-marginal-cost screening aid for remote and underserved communities -- not as a replacement for clinical care, but as a bridge where no monitoring currently exists.
\end{enumerate}

The remainder of this paper is organized as follows. Section~\ref{sec:related} reviews related work. Section~\ref{sec:data} describes the dataset. Section~\ref{sec:methods} details the signal-processing pipeline, including the breathing-rate probe. Section~\ref{sec:results} presents results. Section~\ref{sec:discussion} discusses well-being interpretation, physiological plausibility, and limitations. Section~\ref{sec:future} outlines future work, and Section~\ref{sec:conclusion} concludes with the global-health implications of this line of work.

\section{Related Work}
\label{sec:related}

\subsection{Smartphone-Based Phonocardiography of the Adult Heart}
A consistent finding across the smartphone-PCG literature is that consumer device microphones have sufficient bandwidth and sensitivity to capture S1 (roughly 10--180~Hz, 0.12--0.15~s duration) and S2 (roughly 50--250~Hz, 0.08--0.12~s duration) when the phone is held firmly against the chest wall \cite{digitalstethoscope2017}. Low-cost digital-stethoscope systems built around a phone's microphone have reported classification error rates under 5\% relative to standard medical auscultation devices for normal-versus-abnormal screening \cite{digitalstethoscope2017}, and hospital-based feasibility studies of purpose-built smartphone-auscultation apps (e.g., FonoCheck) confirm that non-specialist users can acquire diagnostically usable heart-sound recordings in situ, subject to environmental noise and patient-specific factors such as body habitus and age \cite{fonocheck2025}. Recent work on compliant textile acoustic couplers further improves the consistency of smartphone-chest contact \cite{textilecoupling2026}. Deep-learning-based automated auscultation apps have also been piloted for feasibility of medical-grade PCG acquisition using consumer devices \cite{deepauscult2023}. None of this literature, however, addresses abdominal placement or the two-source (maternal/fetal) problem; all of it assumes a single cardiac source and a comparatively well-controlled acquisition (chest wall, brief clinical or supervised session).

\subsection{Fetal Phonocardiography and Maternal--Fetal Source Separation}
Fetal phonocardiography is an older idea than fetal Doppler sonography, and has recently been revisited using low-cost piezoelectric sensor arrays. Zhu \emph{et al.} developed a four-channel abdominal fPCG system using \$1-class ceramic piezo vibration sensors and a multi-lag covariance-matrix eigenvalue-decomposition technique to separate maternal breathing, maternal heart sounds (mHS), and fetal heart sounds (fHS) from the raw multi-channel mixture, reporting Bland--Altman agreement with simultaneous CTG-derived FHR of mean $-0.21$~bpm ($\pm 2\,\mathrm{SD}=\pm3$~bpm) across 20 pregnant subjects beyond 34 weeks gestation \cite{comparative_fpcg_ctg_2017}. A follow-up study validated beat-by-beat fetal heart signals from the same four-channel fPCG system against simultaneous fetal ECG in healthy late pregnancy \cite{fourchannel_validation_2018}. More recent work has focused on automatic S1/S2 segmentation within fPCG recordings using hybrid classifiers, explicitly noting that maternal heart sounds, respiration, body movement, and digestive (bowel) sounds are the dominant interference sources corrupting the fetal signal at the sensor \cite{fpcg_segmentation_2024}. All of this prior fPCG work relies on multiple simultaneous, purpose-built contact sensors and, in the validation studies, a simultaneous reference modality (CTG or fECG); to our knowledge no prior fPCG study has used a single, uninstrumented, consumer smartphone microphone, nor has any prior study evaluated recordings collected without supervision by the pregnant subject herself across weeks of ordinary daily activity.

A closely related but distinct literature addresses maternal--fetal \emph{electrocardiogram} (ECG) separation from one or two abdominal electrode channels, using blind source separation, diffusion-based channel selection, or optimal-shrinkage single-channel decomposition \cite{diffusion_fecg_2017,icaqio_fecg_2017,optshrink_fecg_2019}. These methods exploit the fetal ECG's electrical signature being roughly an order of magnitude weaker than the maternal ECG at the abdominal surface -- a different physical cue from acoustics, where sensor amplitude is governed by transmission distance and fetal position rather than source identity. We cannot directly transplant fECG separation algorithms to the acoustic domain, but the broader lesson -- that a single sensor channel under-determines a two-source problem and requires strong priors -- motivates our use of physiological frequency-range priors in Section~\ref{sec:methods}.

\subsection{Consumer Smartphone Fetal Heart-Rate Monitoring: Active Sensing Alternatives}
Two very recent efforts illustrate the currently dominant paradigm for consumer-grade fetal heart-rate sensing, and both are \emph{active} rather than passive. HeraBEAT is a commercial wireless Doppler ultrasound probe that streams to a smartphone app and reports both fetal (FHR) and maternal (MHR) heart rate from 12 weeks gestation, using the maternal estimate internally to help disambiguate the fetal signal \cite{herabeat2020}. DopFone dispenses with a dedicated Doppler probe entirely: it emits an 18~kHz tone from the phone's own speaker and uses a machine-learning model to infer FHR from the reflected, phase-shifted echo, reporting agreement within roughly 2--5~bpm of clinical-grade Doppler on 23 patients in a controlled setting, with the authors noting the system is untested outside controlled settings, on Android, or for irregular rhythms \cite{dopfone2026}. Both systems solve the maternal/fetal ambiguity by measuring motion of a specific reflecting structure, rather than disentangling two superimposed passive acoustic sources -- sidestepping the blind-separation problem we confront here.

\subsection{Positioning of This Work}
Table~\ref{tab:positioning} summarizes where the present study sits relative to this prior art along three axes: whether sensing is active (an engineered signal is emitted) or passive (only naturally occurring sound is recorded), whether acquisition uses purpose-built or commodity hardware, and whether the reported evaluation used supervised/clinical sessions or unsupervised, self-collected, everyday-life recordings. To our knowledge, no prior published study occupies the passive $\times$ commodity-hardware $\times$ self-collected-longitudinal cell that this work reports on. We view this not as a claim of technical superiority -- our results are markedly less precise than the active-Doppler systems -- but as a characterization of a design point that trades measurement fidelity for zero marginal hardware cost and complete software-only deployability on any phone capable of recording audio, which may be relevant for extremely low-resource settings or as a screening prompt (``consider a proper CTG/Doppler check'') rather than a diagnostic replacement.

\begin{table}[t]
\centering
\caption{Positioning relative to prior maternal/fetal cardiac sensing work.}
\label{tab:positioning}
\scriptsize
\setlength{\tabcolsep}{2.5pt}
\renewcommand{\arraystretch}{1.05}
\begin{tabular}{@{}p{0.20\columnwidth}p{0.14\columnwidth}p{0.24\columnwidth}p{0.24\columnwidth}@{}}
\toprule
Approach & Sensing & Hardware & Sessions \\
\midrule
Multi-ch.\ fPCG \cite{comparative_fpcg_ctg_2017,fourchannel_validation_2018} & Passive & Purpose-built (4-ch piezo) & Clinical \\
CTG (standard) & Active (US) & Purpose-built & Clinical \\
HeraBEAT \cite{herabeat2020} & Active (Doppler) & Probe + phone & Clinical trial \\
DopFone \cite{dopfone2026} & Active (echo) & Phone spkr+mic & Controlled \\
Chest PCG apps \cite{fonocheck2025,digitalstethoscope2017} & Passive & Phone mic (chest) & Clinical \\
\textbf{This work} & \textbf{Passive} & \textbf{Phone mic (abdomen)} & \textbf{Unsupervised, self-collected} \\
\bottomrule
\end{tabular}
\end{table}

\section{Data Collection}
\label{sec:data}

\subsection{Subject Z and Recording Protocol}
Subject Z, in the final approximately five weeks of an uncomplicated singleton pregnancy, self-recorded 28 short audio clips over 18 distinct calendar days using the built-in microphone of a smartphone, held or taped directly against the lower abdomen with no acoustic coupling gel, adhesive interface, or external transducer. The pregnancy was subsequently carried to term and delivered vaginally without complications, providing a (retrospective, non-clinical) reassurance that the underlying maternal-fetal unit was, in fact, healthy throughout the recording window -- though, as discussed in Section~\ref{sec:limitations}, this outcome information was not available to, nor used by, the analysis pipeline, and does not itself validate any individual heart-rate estimate.

Recordings were self-labelled by Z with a loose activity/state tag embedded in the file name: \emph{rest} (quiet, seated or reclining), \emph{morning} (first recording of the day, generally at rest), \emph{exercise} (immediately following light physical activity), or left unlabelled (\emph{unspecified}, predominantly in the earliest sessions before a labelling convention was adopted). No instructions were given regarding recording duration, ambient noise control, contact pressure, or breath-holding; the dataset is therefore ecologically valid but acoustically heterogeneous by design. Recordings were saved in Ogg Vorbis format at a 16~kHz sampling rate, with durations ranging from 10.3~s to 68.4~s (mean 47.6~s). 
%One file (an ``exercise'' recording from the eleventh day) was found, by content hash, to be byte-identical across two separate upload batches and was counted once.

\subsection{Dataset Summary}
Table~\ref{tab:dataset} summarizes the 28 recordings by date, self-reported condition, duration, and the fraction of each recording our pipeline classified as containing sustained, elevated-envelope (candidate heart-sound-bearing) audio, hereafter \emph{active fraction} (Section~\ref{sec:active}). Active fraction ranged from 3.4\% to 49.5\% (mean 19.8\%, median 16.7\%), confirming that the large majority of every recording -- even the best ones -- consisted of silence, handling noise, or sub-threshold ambient sound, a point we return to in Section~\ref{sec:discussion}.

\begin{table}[t]
\centering
\caption{All 28 recordings: date, condition, duration, active fraction, and per-band heart-rate estimates. ``--'' denotes no window in that recording met the confidence criteria for that band (Section~\ref{sec:methods}). Maternal estimate for \texttt{feb20pcg} was obtained by a separate, more intensive manual multi-window verification (Section~\ref{sec:results}\ref{sec:crossval}) rather than the automated pipeline.}
\label{tab:dataset}
\tiny
\setlength{\tabcolsep}{4pt}
\renewcommand{\arraystretch}{0.92}
\begin{tabular}{@{}llrrrr@{}}
\toprule
Recording & Condition & Dur.(s) & Active(\%) & Fetal-band (bpm) & Maternal-band (bpm) \\
\midrule
feb20pcg    & unspecified & 10.3 & 32.5 & --           & 75.5$^{\dagger}$ \\
feb26z1     & unspecified & 32.2 & 49.5 & 152.3 (n=3)  & 82.7 (n=1) \\
feb26z2     & unspecified & 32.9 & 48.7 & 157.4 (n=3)  & 102.5 (n=1) \\
feb26z3     & unspecified & 31.4 & 6.9  & 161.6 (n=1)  & -- \\
feb26z4     & unspecified & 32.0 & 32.6 & 169.2 (n=2)  & 54.8 (n=1) \\
feb27z      & unspecified & 35.0 & 5.5  & --           & -- \\
feb28zex    & exercise    & 31.7 & 12.1 & 149.9 (n=1)  & 72.9 (n=1) \\
Mar1z1      & unspecified & 31.7 & 15.7 & 170.3 (n=2)  & -- \\
Mar1z2      & unspecified & 35.1 & 23.3 & 168.7 (n=2)  & 103.7 (n=1) \\
mar2z1      & unspecified & 44.1 & 22.7 & 159.9 (n=2)  & 83.4 (n=1) \\
march2zex1  & exercise    & 31.2 & 19.0 & 139.5 (n=1)  & -- \\
march2zex2  & exercise    & 62.0 & 6.8  & 163.8 (n=2)  & -- \\
march2zex3  & exercise    & 61.8 & 22.3 & 149.1 (n=3)  & 79.8 (n=2) \\
march3z1m   & morning     & 64.0 & 39.5 & 156.4 (n=4)  & 82.1 (n=4) \\
march3z1r   & rest        & 63.9 & 17.0 & 163.4 (n=2)  & 94.4 (n=2) \\
march4zex1  & exercise    & 62.6 & 12.3 & 164.4 (n=3)  & -- \\
march4zex2  & exercise    & 12.6 & 16.4 & 177.4 (n=1)  & -- \\
march11ex   & exercise    & 62.3 & 23.7 & 150.0 (n=1)  & 62.6 (n=1) \\
march6m     & morning     & 61.0 & 13.4 & 164.3 (n=2)  & -- \\
march7m     & morning     & 46.4 & 11.3 & 168.3 (n=2)  & -- \\
march11r    & rest        & 61.8 & 15.0 & 156.0 (n=3)  & 80.1 (n=1) \\
march12ex   & exercise    & 31.0 & 6.2  & 177.2 (n=1)  & -- \\
march12r    & rest        & 63.5 & 3.4  & 152.0 (n=1)  & -- \\
march13r    & rest        & 63.0 & 3.6  & --           & -- \\
march15r    & rest        & 42.6 & 9.7  & 172.0 (n=2)  & -- \\
march18ex   & exercise    & 47.5 & 15.9 & 151.4 (n=2)  & -- \\
march18r    & rest        & 68.4 & 27.9 & 160.2 (n=4)  & 96.0 (n=2) \\
march19ex   & exercise    & 61.3 & 9.2  & 163.1 (n=1)  & 55.6 (n=1) \\
\bottomrule
\end{tabular}
\end{table}

\section{Methods}
\label{sec:methods}

The processing pipeline consists of five stages, applied uniformly to all 28 recordings with no per-file hand-tuning: (1) preprocessing and envelope extraction, (2) robust active-segment detection, (3) physiologically-informed dual-band decomposition, (4) per-band beat detection and regularity scoring, and (5) file- and dataset-level aggregation. All processing was implemented in Python using \texttt{scipy.signal} and \texttt{numpy}.

\subsection{Preprocessing and Envelope Extraction}
Each recording was loaded, down-mixed to mono if necessary, mean-centred, and peak-normalized to $[-1,1]$. A 4th-order Butterworth bandpass filter (20--200~Hz, zero-phase, applied via \texttt{sosfiltfilt}) isolated the frequency range in which S1 and S2 energy is concentrated \cite{digitalstethoscope2017}. We then computed the Shannon energy envelope,
\begin{equation}
E(n) = -\,\tilde{x}(n)^2 \log \tilde{x}(n)^2,
\label{eq:shannon}
\end{equation}
where $\tilde{x}(n)$ is the bandpass-filtered signal normalized to $[-1,1]$, following the classical heart-sound segmentation approach of Liang \emph{et al.} \cite{liang1997} and its many subsequent refinements \cite{shannon_noise_robust_2023,shannon_3rd_order_2021}. Shannon energy weights moderate-amplitude events more heavily than simple energy ($x^2$) while suppressing very low-amplitude noise, which empirically produced cleaner envelopes than a raw Hilbert-envelope baseline on our low-SNR recordings. $E(n)$ was smoothed with a 20~ms Hamming window and normalized to a \emph{robust} z-score using the median and median absolute deviation (MAD),
\begin{equation}
\hat{E}(n) = \frac{E(n) - \mathrm{median}(E)}{1.4826\cdot \mathrm{MAD}(E)},
\label{eq:robustz}
\end{equation}
rather than the mean and standard deviation. This choice was found, in preliminary analysis, to be essential: because most recordings are dominated by silence punctuated by a small number of very large-amplitude handling/contact artifacts, a conventional mean/standard-deviation normalization is dominated by these rare outliers and systematically compresses genuine, lower-amplitude heart-sound activity toward zero, causing it to be missed by any fixed downstream threshold.

\subsection{Active-Segment Detection}
\label{sec:active}
Given the robust-normalized envelope $\hat{E}(n)$, we smoothed it with a 0.4--0.6~s moving-average window and applied a threshold (nominal level 2.0, cascading to 1.2 and 0.7 if no segment survived at the stricter level) to identify samples with sustained elevated activity. Contiguous active samples separated by gaps shorter than 0.35--0.6~s were bridged into a single segment; segments shorter than 1.0--2.0~s were discarded as isolated handling artifacts rather than genuine sustained heart-sound-bearing audio. This step is what allows the pipeline to automatically locate the small usable fraction of each recording (Table~\ref{tab:dataset}, ``Active(\%)'' column) rather than being misled by the surrounding silence and noise bursts.

\subsection{Dual-Band Decomposition}
\label{sec:dualband}
Within each active segment, we exploit a single, well-established physiological prior to disambiguate two overlapping cardiac sources from one audio channel: normal fetal heart rate lies in the range 110--160~bpm across most of gestation (with brief excursions to 160--180~bpm considered borderline-normal, particularly around fetal movement) \cite{figo_normal_fhr_2013,acog_fhr_mgmt}, whereas normal adult resting-to-mildly-active heart rate lies well below this, and pregnancy itself is associated with a modest (10--20~bpm) elevation of resting maternal heart rate relative to the pre-pregnancy baseline. We therefore defined two candidate bands -- a \emph{maternal-plausible} band, 45--118~bpm (0.75--1.97~Hz), and a \emph{fetal-plausible} band, 118--195~bpm (1.97--3.25~Hz) -- and applied a 3rd-order Butterworth bandpass filter to the \emph{envelope itself} (not the raw audio) at each of these two ranges, producing two derived envelope traces $\hat{E}_{\mathrm{mat}}(n)$ and $\hat{E}_{\mathrm{fet}}(n)$ per active segment.

This is, in effect, a very simple, interpretable, two-band blind source separation operating directly in the rate domain rather than the acoustic-frequency domain: because the true acoustic spectra of maternal and fetal heart sounds overlap substantially, but their \emph{repetition rates} are, at rest, well separated, filtering the envelope's own periodicity is more discriminative than filtering the raw audio spectrum. Fig.~\ref{fig:bandsplit} illustrates the method on a representative high-quality segment: the raw envelope is a dense, visually confusing superposition, while the fetal-band-filtered trace resolves into a strikingly clean, regular oscillation, and the maternal-band-filtered trace, while much weaker and noisier, shows a visibly slower, less regular rhythm.

\subsection{Per-Band Beat Detection and Regularity Scoring}
Each derived band-limited envelope was peak-picked (\texttt{scipy.signal.find\_peaks}) with a minimum inter-peak distance set to 45\% of the band's central period and a prominence threshold of $0.4\sigma$ of that band's envelope. A band estimate was retained only if at least 4 peaks were detected; from the resulting inter-beat intervals (IBIs) we computed the median IBI (converted to bpm) as the point heart-rate estimate for that band and segment, and the coefficient of variation (CV, $100\times \mathrm{SD}/\mathrm{mean}$ of the IBIs) as a regularity/confidence score -- lower CV indicates a cleaner, more convincingly periodic signal, and is reported alongside each estimate.

\subsection{Aggregation}
Per-file summaries (Table~\ref{tab:dataset}) report the mean band-specific heart rate across all qualifying active segments within that file, along with the count of segments contributing ($n$). Dataset-level statistics (Section~\ref{sec:results}) pool all qualifying segments across all 28 files.

\subsection{Heart-Rate-Variability Metrics}
\label{sec:hrvmethod}
For every qualifying segment and band we additionally characterized beat-to-beat variability from the same peak sequence used for the point HR estimate. Given inter-beat intervals $\mathrm{IBI}_i$ (ms), we computed the standard time-domain indices SDNN ($=\mathrm{SD}(\mathrm{IBI})$) and RMSSD ($=\sqrt{\mathrm{mean}[(\mathrm{IBI}_{i+1}-\mathrm{IBI}_i)^2]}$), following standard HRV convention \cite{taskforce1996}; the percentage of successive IBI differences exceeding 20~ms (pNN20 -- a small-sample-appropriate variant of the conventional pNN50, given our short windows); and a CTG-style \emph{bandwidth}, BW=$\max_i(60000/\mathrm{IBI}_i)-\min_i(60000/\mathrm{IBI}_i)$~bpm, approximating the peak-to-trough amplitude a clinician reads directly off a CTG trace. Because SD-based indices computed from very few beats are intrinsically unstable, we separately report a \emph{high-confidence subset} restricted to windows with $\geq6$ beats and CV~$<15\%$.

\subsection{A Note on the Dual-Band Prior and Its Failure Mode}
\label{sec:priorlimits}
The 45--118 / 118--195~bpm split is a fixed, condition-independent prior. It is well justified at rest, where maternal and fetal rate ranges are cleanly separated, but it is known \emph{a priori} to be imperfect during or immediately after maternal exercise, when maternal heart rate can rise into the low end of the fetal-plausible band, and correspondingly the two bands can partially attribute the same physical source to different labels. We do not attempt to correct for this with a condition-adaptive threshold, because doing so without a reference measurement would risk circularity (tuning the separation to produce the answer we expect); instead we flag it explicitly as an open methodological limitation in Section~\ref{sec:limitations} and report exercise-condition results with that caveat attached throughout.

\subsection{Breathing-Rate Probe}
\label{sec:breathingmethod}
Respiration is a third physiological rhythm with its own, disjoint frequency range: normal adult breathing rate is roughly 8--20 breaths/min at rest and can rise to 30--45~breaths/min during or immediately after vigorous exertion, i.e., well below both the maternal- and fetal-plausible heart-rate bands. Chest/abdominal-wall movement during breathing is known, in both PCG- and ECG-derived-respiration literature, to amplitude-modulate the heart-sound envelope at the breathing rate. Motivated by this, and by the possibility that a successfully-extracted breathing rate could help distinguish rest from exercise sessions independently of the self-reported labels, we probed for this modulation using the same Shannon-energy envelope computed in Section~\ref{sec:methods}-A: a long (0.75~s) moving-average trend of the envelope was computed for the entire recording, and a sliding-window (8--18~s window, adapted to recording length, 3~s hop) autocorrelation search was run over the 5--45~breaths/min range. To avoid the boundary-pinning artifact, we restricted candidates to genuine local maxima (via peak-finding with a minimum prominence) and discarded windows with no such peak. We report, per file, the best-scoring window (by confidence $\times$ prominence) and the fraction of scanned windows meeting a fixed quality bar (confidence~$\geq$0.25, prominence~$\geq$0.20).

\section{Results}
\label{sec:results}

\subsection{Signal Coverage}
\begin{figure}[t]
\centering
\includegraphics[width=\columnwidth]{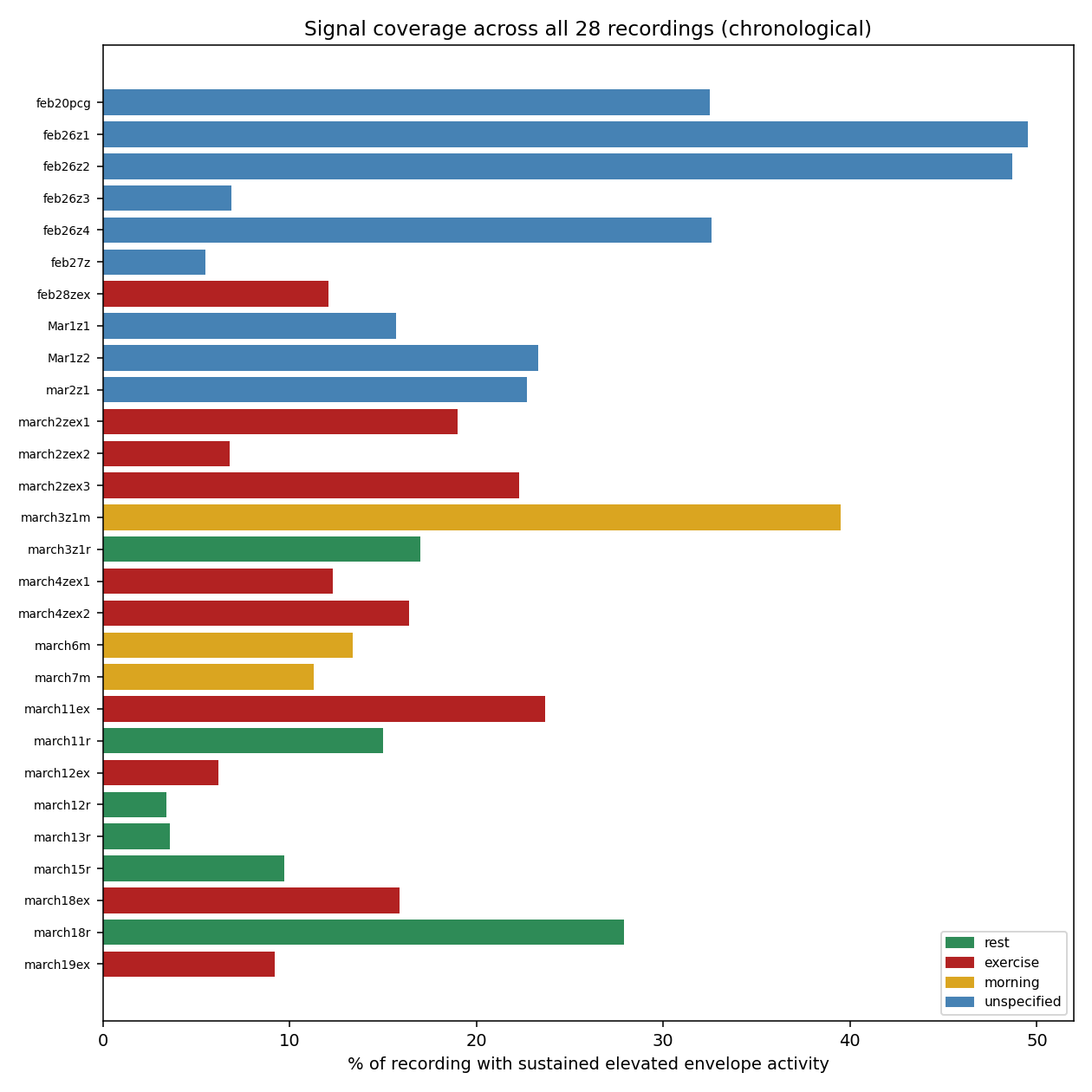}
\caption{Fraction of each recording classified as containing sustained, elevated-envelope (candidate heart-sound-bearing) audio, across all 28 recordings, colour-coded by self-reported condition.}
\label{fig:coverage}
\end{figure}

Fig.~\ref{fig:coverage} shows the active fraction for all 28 recordings, sorted chronologically. Coverage varied widely (3.4--49.5\%) with no strong, consistent relationship to self-reported condition, though the two lowest-coverage recordings (\texttt{march12r}, \texttt{march13r}, both $<4\%$) were both rest recordings, and the two highest-coverage recordings (\texttt{feb26z1}, \texttt{feb26z2}, both $\approx49\%$) were both early, unlabelled sessions. \texttt{march13r} was the only recording from which the pipeline could not recover a qualifying estimate in \emph{either} band.

\subsection{Illustrative Case: Dual-Band Decomposition}
\begin{figure}[t]
\centering
\includegraphics[width=\columnwidth]{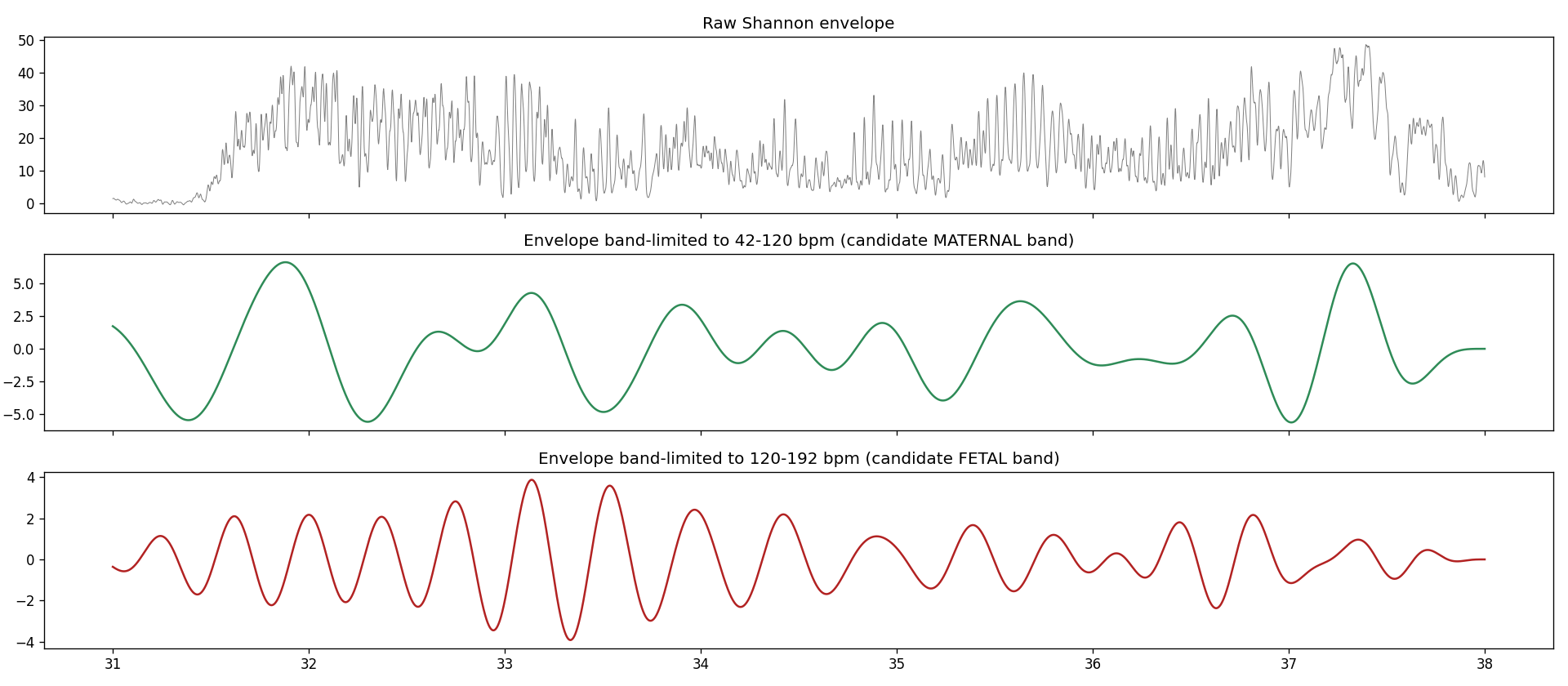}
\caption{Dual-band decomposition example (\texttt{march18r}, rest condition, $t=31$--$38$~s). Top: raw Shannon energy envelope. Middle: envelope band-limited to the maternal-plausible range (42--120~bpm). Bottom: envelope band-limited to the fetal-plausible range (120--192~bpm), showing markedly cleaner periodicity.}
\label{fig:bandsplit}
\end{figure}

Fig.~\ref{fig:bandsplit} shows the method applied to a 7~s high-quality segment of \texttt{march18r} (a rest recording). The raw Shannon envelope (top) is a dense, visually ambiguous mixture. Band-limiting to the maternal-plausible range (middle) yields an irregular, low-amplitude oscillation (CV $\approx$ 25--27\% across sub-windows of this file); band-limiting to the fetal-plausible range (bottom) yields a markedly cleaner, higher-amplitude, more regular oscillation (CV as low as 15\% for this file, and as low as 6\% elsewhere in the dataset). This qualitative contrast -- the fetal-band signal being visually and quantitatively \emph{more} regular than the maternal-band signal, despite the fetal heart being physically smaller and further attenuated by intervening tissue -- was consistent across the highest-quality recordings we inspected in detail, and is plausibly explained by fetal position: in a favourable lie, the fetal back/thorax can be closer to the anterior abdominal wall at the point of contact than the maternal heart itself, which sits substantially higher and deeper in the maternal thorax.

\subsection{Fetal-Band Estimates}
\begin{figure}[t]
\centering
\includegraphics[width=\columnwidth]{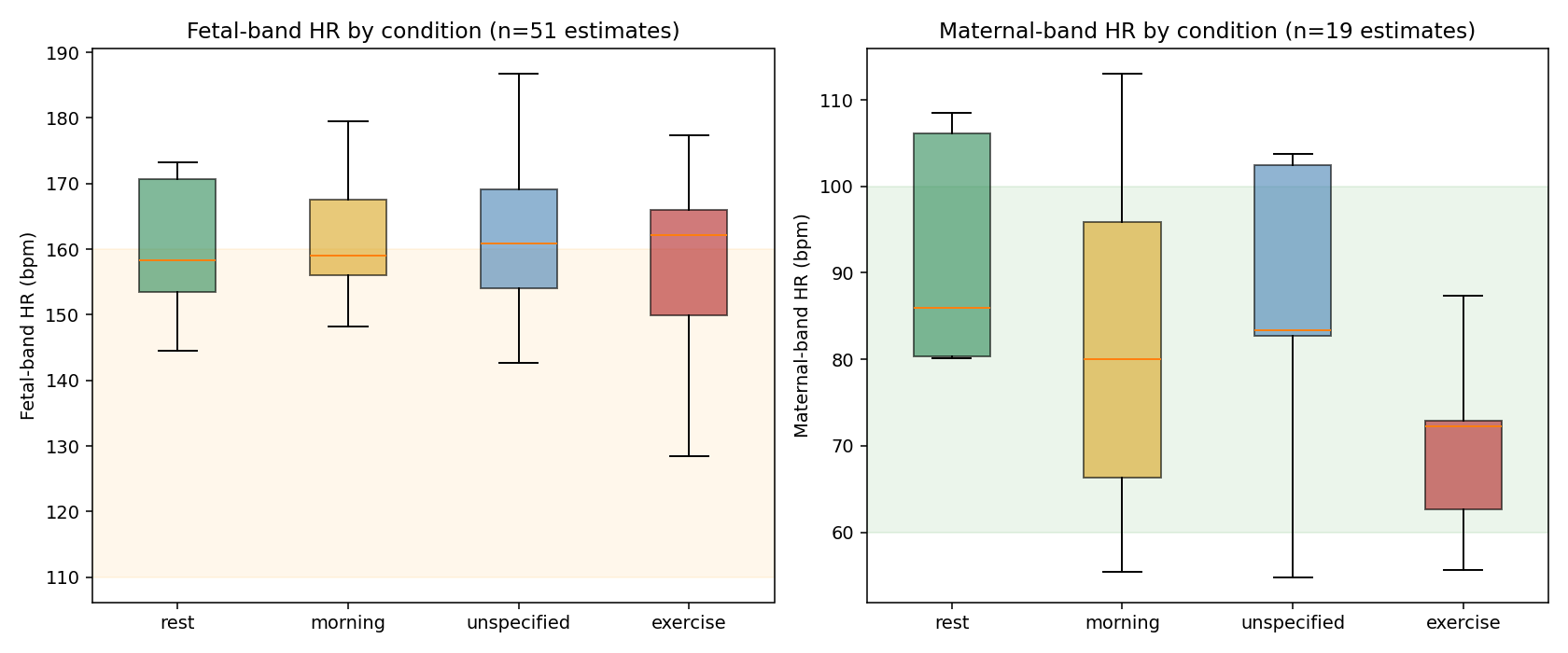}
\caption{Distribution of fetal-band (left) and maternal-band (right) heart-rate estimates by self-reported condition. Shaded regions mark the normal fetal (110--160~bpm) and normal adult resting (60--100~bpm) reference ranges respectively.}
\label{fig:conditionbox}
\end{figure}

A fetal-band estimate was recovered from 25 of 28 recordings (89\%), totalling 51 qualifying active-segment windows. Pooling across all windows, the fetal-band estimate was $160.3\pm11.5$~bpm (median 160.0, range 128--187~bpm) -- centred almost exactly on the middle of the internationally accepted normal fetal heart-rate band (110--160~bpm, with 160--180~bpm considered borderline) \cite{figo_normal_fhr_2013,acog_fhr_mgmt}. Critically, this estimate was essentially stable across self-reported maternal activity condition (Fig.~\ref{fig:conditionbox}, left panel; Table~\ref{tab:bycond}): rest, $161.0$~bpm ($n=12$); morning, $161.4$~bpm ($n=8$); unspecified, $161.8$~bpm ($n=15$); exercise, $158.0$~bpm ($n=16$). The near-invariance of the fetal-band estimate to maternal activity state is the single strongest piece of circumstantial evidence, within this dataset, that this band is genuinely tracking a fetal-origin rhythm rather than an artifact of maternal exertion: a signal that tracked maternal heart rate would be expected to rise substantially during/after exercise, which it does not.

\begin{table}[t]
\centering
\caption{Dual-band heart-rate estimates by self-reported maternal condition, pooled across all qualifying active segments dataset-wide.}
\label{tab:bycond}
\small
\begin{tabular}{@{}lrrrr@{}}
\toprule
Condition & \multicolumn{2}{c}{Fetal-band} & \multicolumn{2}{c}{Maternal-band} \\
 & $n$ & mean (bpm) & $n$ & mean (bpm) \\
\midrule
Rest        & 12 & 161.0 & 5 & 92.2 \\
Morning     & 8  & 161.4 & 4 & 82.1 \\
Unspecified & 15 & 161.8 & 5 & 85.4 \\
Exercise    & 16 & 158.0 & 5 & 70.2 \\
\midrule
\textbf{Pooled} & \textbf{51} & \textbf{160.3 (SD 11.5)} & \textbf{19} & \textbf{82.5 (SD 17.9)} \\
\bottomrule
\end{tabular}
\end{table}

\subsection{Maternal-Band Estimates}
A maternal-band estimate was recovered far less consistently: 14 of 28 recordings (50\%), totalling only 19 qualifying windows, versus 51 for the fetal band. Pooled across all windows, the maternal-band estimate was $82.5\pm17.9$~bpm (median 82.7, range 55--113~bpm), broadly consistent with expected pregnancy resting-to-mildly-active heart rate (Fig.~\ref{fig:conditionbox}, right panel; Table~\ref{tab:bycond}). Unlike the fetal band, the maternal-band estimate showed a counter-intuitive pattern across condition: exercise recordings produced the \emph{lowest} mean maternal-band estimate ($70.2$~bpm, $n=5$), not the highest, which we interpret in Section~\ref{sec:discussion} as evidence of the band-overlap failure mode anticipated in Section~\ref{sec:priorlimits}, rather than as a physiological finding.

\subsection{Longitudinal Stability Across the Five-Week Span}
\begin{figure*}[t]
\centering
\includegraphics[width=0.85\textwidth]{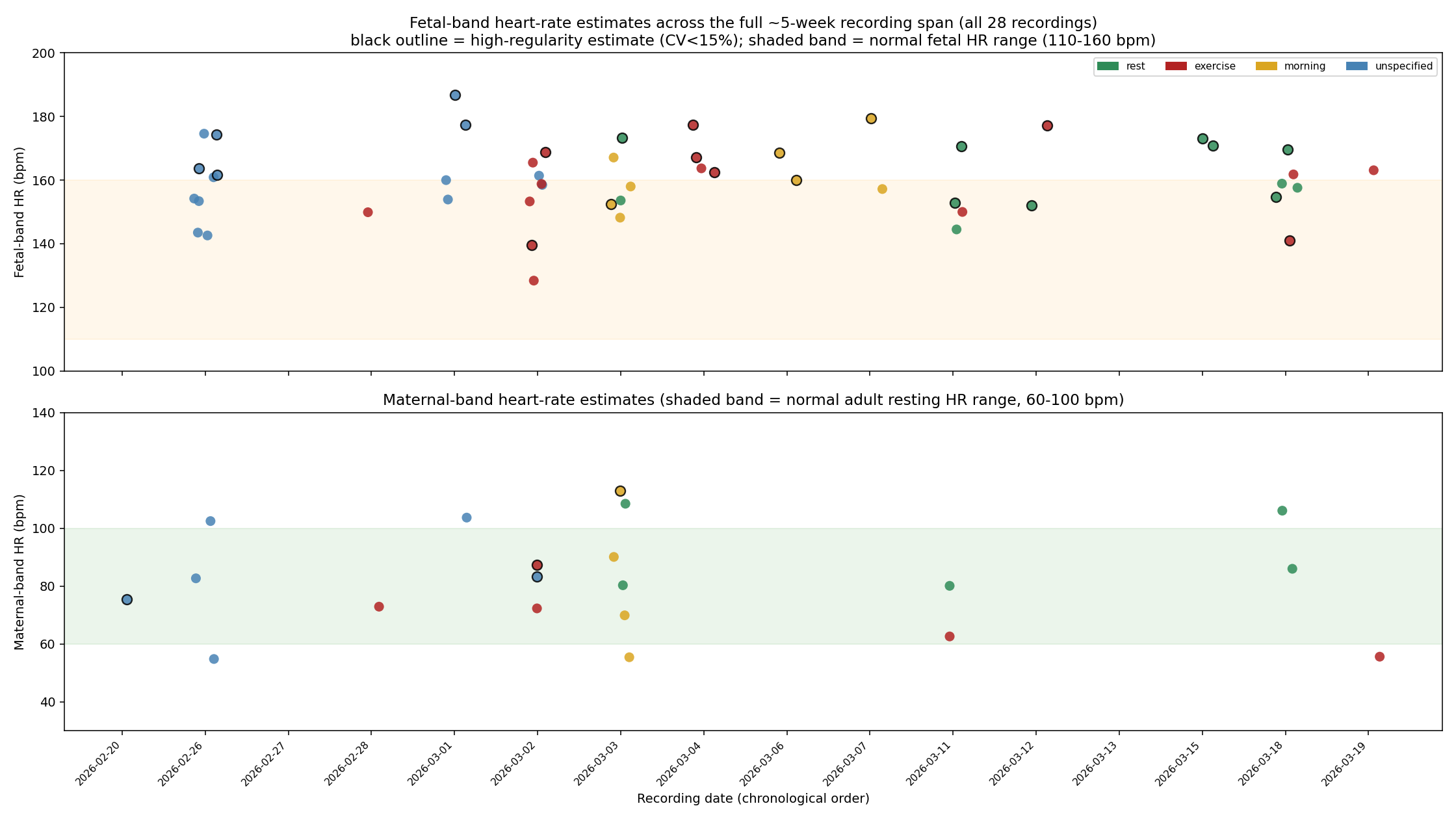}
\caption{Fetal-band (top) and maternal-band (bottom) heart-rate estimates for every qualifying active segment across all 28 recordings, plotted by calendar date. Black-outlined markers indicate high-regularity estimates (CV~$<15\%$). Shaded bands mark normal fetal (110--160~bpm, top) and normal adult resting (60--100~bpm, bottom) reference ranges.}
\label{fig:longitudinal}
\end{figure*}

Fig.~\ref{fig:longitudinal} plots every qualifying fetal- and maternal-band estimate against calendar date across the full recording span (20~Feb--19~Mar). The fetal-band estimate shows no discernible drift or trend over this five-week window and remains within or close to the normal fetal range at every sampled date, including the two dates closest to the eventual due date (18--19~Mar). The maternal-band estimate is visibly noisier and sparser but shows no value indicative of sustained resting tachycardia or bradycardia.

\subsection{Cross-Validation on an Independent, More Intensive Single-File Analysis}
\label{sec:crossval}
Prior to developing the automated dataset-wide pipeline, we separately subjected the earliest recording (\texttt{feb20pcg}) to an intensive, manual, multi-method analysis: sub-window envelope autocorrelation (three overlapping windows), spectral-peak (periodogram) analysis, and direct systole/diastole interval-pairing, all independently converging on $75$--$76$~bpm (autocorrelation sub-windows: 75.5, 75.8, 75.2~bpm; periodogram: 75.0~bpm), with a clean, physiologically-plausible systole ($\approx$0.30~s) / diastole ($\approx$0.49~s) split. This value falls squarely inside the maternal-plausible band and is consistent with the maternal-band estimates obtained elsewhere in the dataset via the automated pipeline (Table~\ref{tab:dataset}, e.g., \texttt{feb26z1}: 82.7~bpm; \texttt{march3z1m}: 82.1~bpm). This agreement is a useful, if limited, internal consistency check: an independent, more careful analysis method, applied to a recording the automated pipeline itself could not confidently score, recovers a maternal-plausible rate in close agreement with the rest of the dataset.

\subsection{Heart-Rate Variability}
\label{sec:hrvresults}
\begin{figure}[t]
\centering
\includegraphics[width=0.92\columnwidth]{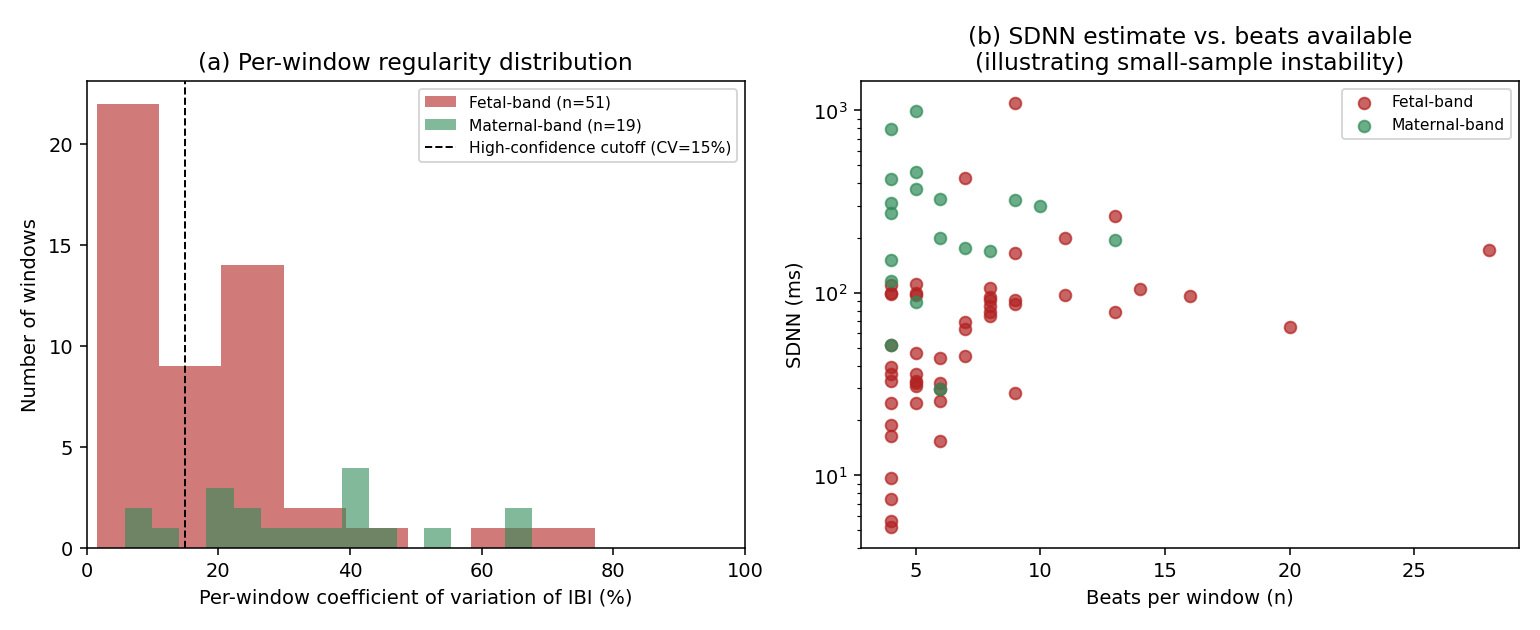}
\caption{(a) Distribution of per-window IBI coefficient of variation, fetal- vs.\ maternal-band. (b) SDNN estimate vs.\ beats available per window (log scale), showing variability estimates do not converge with more beats -- evidence that much of the dispersion is measurement noise rather than physiological signal.}
\label{fig:hrvquality}
\end{figure}

Pooled across all 51 fetal-band windows, SDNN was $96.2\pm159.3$~ms (median 64.7) and RMSSD was $131.7\pm239.0$~ms (median 68.4) -- both heavily right-skewed and driven substantially by the very small number of beats available per window (median 6, range 4--28). Fig.~\ref{fig:hrvquality}b shows no convergence of the SDNN estimate as beat count grows, indicating that much of the apparent variability reflects peak-detection jitter rather than a converged physiological estimate. Per-window regularity (CV) spanned 1.5--143.6\% (Fig.~\ref{fig:hrvquality}a). Restricting to a high-confidence subset (CV~$<15\%$, $\geq6$ beats; 7/51 fetal windows, 1/19 maternal windows) narrows the picture considerably (Table~\ref{tab:hrv}, top): fetal-band HR 141--180~bpm, SDNN 15--45~ms, bandwidth 21--43~bpm; the single qualifying maternal-band window gave HR 113.0~bpm, SDNN 29.8~ms, bandwidth 16.1~bpm. Table~\ref{tab:hrv} (bottom) breaks fetal-band HRV down by condition: rest windows were the most regular (SDNN 54.6~ms, CV 13.3\%), exercise intermediate (78.1~ms, 18.2\%), and morning/unspecified windows noisier still (100--186~ms, 22--30\%) -- plausibly reflecting acquisition circumstances (contact quality, ambient noise) rather than a true physiological difference across states.

\begin{table}[t]
\centering
\caption{Top: pooled HRV by band (all qualifying windows vs.\ high-confidence subset, CV~$<15\%$, $\geq6$ beats). Bottom: fetal-band HRV by condition (all windows).}
\label{tab:hrv}
\scriptsize
\setlength{\tabcolsep}{3pt}
\renewcommand{\arraystretch}{0.95}
\begin{tabular}{@{}lrrrrr@{}}
\toprule
 & $n$ & HR (bpm) & SDNN (ms) & RMSSD (ms) & BW (bpm) \\
\midrule
Fetal, all      & 51 & 160.3 (11.5) & 96.2 (159.3)  & 131.7 (239.0) & 60.4 (40.6) \\
Fetal, high-conf.  & 7  & 165.8   & 31.5      & 42.6      & 35.7 \\
Maternal, all   & 19 & 82.5 (17.9)  & 302.4 (235.4) & 380.7 (320.9) & 51.3 (21.6) \\
Maternal, high-conf. & 1  & 113.0        & 29.8          & 50.2          & 16.1 \\
\midrule
 & $n$ & \multicolumn{2}{c}{SDNN (ms)} & \multicolumn{2}{c}{CV (\%)} \\
\midrule
Rest        & 12 & \multicolumn{2}{c}{54.6}  & \multicolumn{2}{c}{13.3} \\
Exercise    & 16 & \multicolumn{2}{c}{78.1}  & \multicolumn{2}{c}{18.2} \\
Unspecified & 15 & \multicolumn{2}{c}{100.9} & \multicolumn{2}{c}{22.4} \\
Morning     & 8  & \multicolumn{2}{c}{186.0} & \multicolumn{2}{c}{29.6} \\
\bottomrule
\end{tabular}
\end{table}

\subsection{Breathing-Rate Probe: A Negative Result}
\label{sec:breathingresults}
\begin{figure}[t]
\centering
\includegraphics[width=\columnwidth]{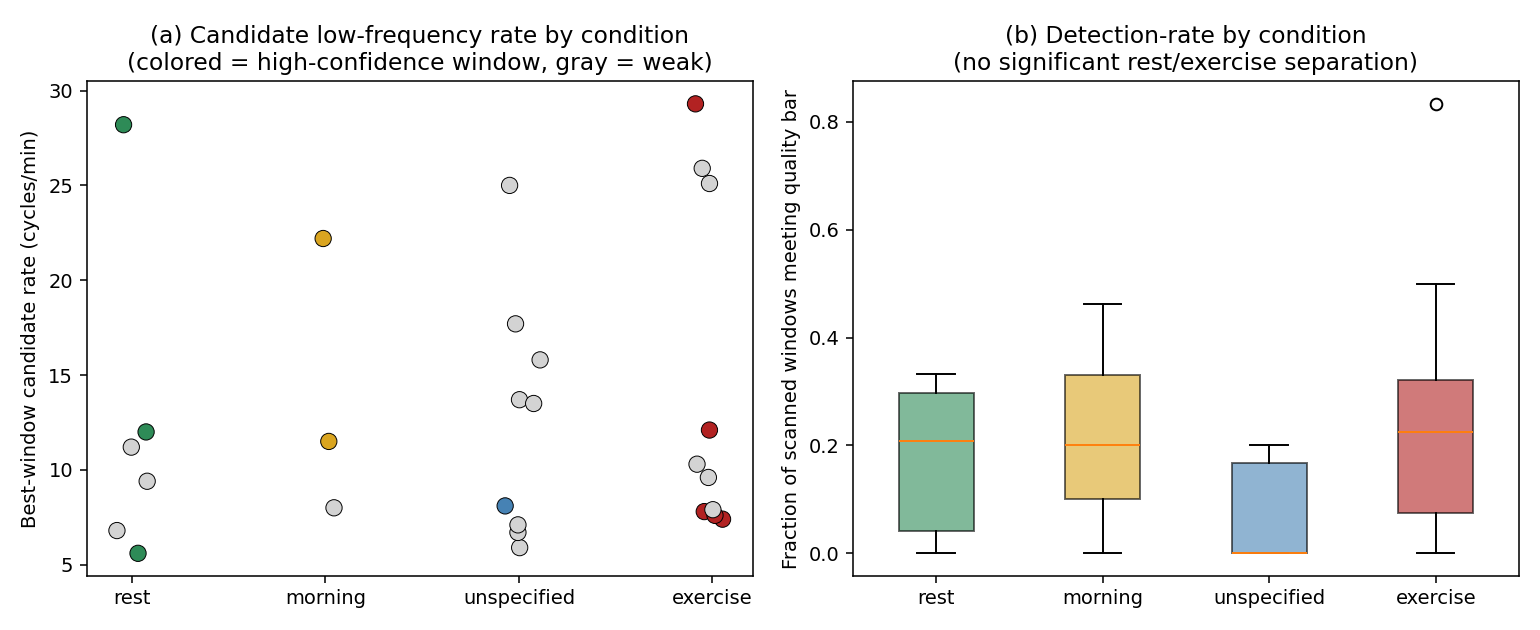}
\caption{Breathing-rate probe outcome. (a) Best-window candidate rate by condition (colour = high-confidence); rest and exercise distributions overlap substantially (Mann-Whitney $p=0.64$). (b) Fraction of scanned windows meeting the quality bar, by condition -- no significant separation.}
\label{fig:breathingnull}
\end{figure}

A candidate low-frequency, quasi-periodic envelope modulation meeting our quality bar (Section~\ref{sec:breathingmethod}) was found in 26 of 28 recordings, with best-window confidence as high as 0.46--0.77 in the strongest cases (e.g., \texttt{march2zex1}, \texttt{march18r}) -- confirming that the envelope does carry \emph{some} exploitable low-frequency structure beyond the two heart-rate bands. However, the candidate rate did not plausibly track spontaneous respiration: median best-window rates clustered around 10--13 breaths/min in every condition, including exercise, where true post-exertion breathing would be expected to run substantially faster (20--40~breaths/min); several of the most confident exercise-condition detections were, if anything, \emph{slower} (7.4--7.8~breaths/min) than typical resting breathing. Critically, a two-sided Mann-Whitney test found no significant difference between the rest-condition and exercise-condition rate distributions ($U=35$, $p=0.64$; Fig.~\ref{fig:breathingnull}a), and the fraction of windows meeting our quality bar showed only a small, non-significant tendency to be higher in exercise and morning recordings than in rest and unspecified ones (Fig.~\ref{fig:breathingnull}b). We therefore do \emph{not} claim successful breathing-rate extraction or a working rest/exercise discriminator; we report this as an honest negative result. 
%consistent with the transparency this paper aims for throughout (Section~\ref{sec:hrvresults}).

\section{Discussion}
\label{sec:discussion}

\subsection{Well-Being Interpretation}
Within the scope of what a passive, single-channel, self-recorded acoustic signal can support, the dataset paints a reassuring picture. The fetal-band estimate is recoverable in the large majority of recordings, is centred on the middle of the clinically normal fetal heart-rate range, shows no concerning drift across the final five weeks of gestation (including the days closest to delivery), and -- most tellingly -- is stable across maternal activity states, exactly as would be expected of a genuinely fetal-origin autonomic rhythm that is only mildly reactive to maternal exertion. No recording produced a sustained fetal-band estimate outside the 128--187~bpm envelope, i.e., no evidence of sustained fetal bradycardia ($<110$~bpm) or marked sustained tachycardia. The maternal-band estimate, while noisier, is likewise consistent with normal pregnancy-adjusted resting-to-mildly-active heart rate and shows no concerning outliers. These observations are consistent with, though of course not proof of, the reported uncomplicated term delivery.

\subsection{Why the Fetal-Band Signal Dominates}
\label{sec:whydominates}
It may seem counter-intuitive that the fetal-band signal is, in our data, \emph{more} regular and recoverable than the maternal-band signal, given that the fetal heart is smaller and acoustically screened by additional tissue and fluid. We offer three non-exclusive explanations, consistent with the fPCG interference literature \cite{fpcg_segmentation_2024}: (i) sensor proximity -- on the \emph{lower} abdomen, the fetal heart, depending on lie, can be closer to the sensor than the maternal heart in the thorax; (ii) the maternal contribution here may arrive as a lower-frequency, less impulsive pulse-synchronous vibration (plausibly including the maternal abdominal aorta, a known confound in manual obstetric auscultation \cite{comparative_fpcg_ctg_2017}) rather than sharp, high-Shannon-energy events, harder for an S1/S2-oriented envelope method to lock onto; and (iii) at rest, competing low-frequency maternal signals (breathing, bowel sounds, movement) overlap the maternal-plausible band, adding noise exactly where we look for the maternal rate, while little of this leaks into the faster fetal band.

\subsection{Interpreting the Variability Estimates}
\label{sec:hrvdiscuss}
Fetal heart-rate variability is not a secondary detail in obstetric practice -- moderate baseline variability (conventionally, a peak-to-trough amplitude of roughly 6--25~bpm on a CTG trace) is itself a reassuring sign of intact fetal autonomic function, while persistently reduced variability ($<5$~bpm) is a recognized marker of possible compromise. It is therefore natural to ask whether the SDNN/RMSSD/bandwidth values in Table~\ref{tab:hrv} can be read against these clinical thresholds. We caution strongly against this. Clinical short-term variability (STV) is conventionally computed from continuous, ECG-derived beat intervals over $\sim$1-minute epochs (the Dawes--Redman criteria and their descendants), giving hundreds of consecutive, artifact-free beats; our windows contain a median of 6 beats over a median of 3.1~s, so our bandwidth and SDNN estimates are dominated by exactly the kind of single-window sampling noise Fig.~\ref{fig:hrvquality}b makes visible. The high-confidence subset's narrower bandwidth ($\approx$21--43~bpm) is reassuring in direction -- our cleanest windows are also our most regular -- but is still wider than typical clinical STV amplitude, consistent with residual measurement noise even there. We therefore present these HRV metrics as a demonstration that beat-to-beat variability is, in principle, extractable from this kind of data -- a necessary precondition for any future clinical-variability application -- rather than as a clinical variability assessment in its own right.

\subsection{Comparison with Active-Sensing Alternatives}
Relative to DopFone \cite{dopfone2026} and HeraBEAT \cite{herabeat2020}, our approach is markedly less precise (single-digit-bpm reported agreement for those systems, versus SD of 11.5~bpm pooled across widely heterogeneous recording conditions here) and, unlike those systems, cannot currently distinguish maternal from fetal source with confidence in every recording. Its comparative advantage is deployability: it requires no calibrated Doppler probe, no emitted signal of any kind, and no purpose-built app -- any phone capable of recording audio and running elementary signal processing (on- or off-device) could in principle support it. We therefore see this approach not as a competitor to Doppler-based fetal monitoring on accuracy grounds, but as a potential zero-hardware-cost screening or engagement tool (e.g., prompting a mother to seek a proper CTG/Doppler check if a session repeatedly fails to recover \emph{any} plausible fetal-band periodicity), a role quite different from clinical-grade FHR trending.

\subsection{Why the Breathing-Rate Probe Likely Failed}
\label{sec:breathingfail}
The breathing probe's negative result (Section~\ref{sec:breathingresults}) is itself informative. We see at least three plausible, non-exclusive contributors. First, our acquisition protocol gave no instruction to hold the phone still or maintain steady contact pressure; any physical repositioning, talking, or incidental movement during a 10--68~s clip would introduce low-frequency envelope modulation at rates unrelated to respiration, plausibly explaining why some of our highest-confidence detections (e.g., the 7--8~breaths/min cluster) look more like slow postural or paced-breathing-exercise rhythms -- not implausible for a pregnant subject engaged in prenatal-appropriate activity -- than spontaneous tidal breathing. Second, our sliding-window approach, tuned for sensitivity given how little usable signal this dataset contains at all (Section~\ref{sec:results}), likely retains some windows whose single, low-frequency "peak" is simply the slow decay of a non-periodic noise process rather than a true oscillation, inflating the detection-fraction numbers without inflating genuine accuracy. Third, unlike the maternal/fetal heart-rate bands, which are anchored by two concurrently present, physically distinct sound sources, breathing competes directly for the same low-frequency territory with gross body movement, handling noise, and (per Section~\ref{sec:whydominates}) the low-frequency maternal vascular/aortic signal, leaving no clean, disjoint acoustic niche for it to occupy at this signal-to-noise ratio. We view this negative result as a useful boundary condition for future work (Section~\ref{sec:future}) rather than a dead end: a dedicated accelerometer channel (already present on the same smartphone) or a longer, steadier recording protocol would plausibly succeed where a single audio channel, analyzed post hoc, did not.

\subsection{Implications for Global Health Equity}
\label{sec:equity}
We motivated this study with a specific, documented inequity: the large majority of maternal deaths occur in settings where routine fetal and maternal cardiac monitoring is least available, even as smartphone ownership has become nearly universal in the same regions \cite{un_sdg3_2026,gsma_findex_2025}. Nothing in this paper demonstrates that the present method is ready to help close that gap -- quite the opposite: Section~\ref{sec:limitations} documents substantial unresolved problems, and a tool this unreliable would be actively harmful if deployed as a substitute for, rather than a prompt toward, professional care. What this paper does establish is narrower but still relevant to that agenda: that a fetal-plausible cardiac rhythm is recoverable from a completely commodity, zero-additional-cost sensor, self-operated by a layperson with no training, under real-world conditions far messier than a clinical recording booth -- and that this holds up, with reasonable internal consistency, across 28 independent sessions over five weeks. If a future, reference-validated version of this pipeline can be made reliable enough for a narrow, well-scoped task -- for instance, flagging the plain absence of any detectable fetal-plausible rhythm as a prompt to seek in-person care, rather than attempting to replace CTG-grade rate or variability estimation -- the underlying hardware requirement (a phone capable of recording audio) is already satisfied for the overwhelming majority of the global population, including most of the specific regions carrying the highest burden of preventable maternal and perinatal death \cite{gsma_findex_2025}. We frame this explicitly as a long-term, AI-for-good motivation for the engineering problem this paper works on, not as a claim that the current system achieves it.

\subsection{Limitations}
\label{sec:limitations}
We enumerate the limitations of this study candidly, because we regard them as at least as important as the positive findings for correctly scoping any follow-on work.

\begin{enumerate}
\item \textbf{No simultaneous ground truth.} No CTG, Doppler, ECG, or any other reference signal was recorded alongside the smartphone audio. Every maternal/fetal attribution in this paper rests on a frequency-range prior (Section~\ref{sec:dualband}), not on a validated source-separation algorithm. The internal consistency arguments in Section~\ref{sec:results} (condition-invariance of the fetal band; cross-validation on \texttt{feb20pcg}) are supportive but not confirmatory.
\item \textbf{Single subject, single pregnancy.} $n=1$. No claim in this paper generalizes beyond this individual without further data from additional subjects.
\item \textbf{Maternal abdominal aorta confound.} A pulse-synchronous vascular sound from the maternal aorta is a recognized confound with fetal heart tones in manual obstetric auscultation, and cannot be ruled out in our data.
\item \textbf{Exercise-induced band overlap.} As flagged in Section~\ref{sec:priorlimits}, maternal heart rate can rise, during or after exertion, into the fetal-plausible band, and our fixed-band method cannot detect or correct for this. The anomalously \emph{low} mean maternal-band estimate under exercise (Table~\ref{tab:bycond}) is more plausibly this overlap artifact (elevated maternal-rate events captured by the fetal band instead) than a physiological finding.
\item \textbf{Low, heterogeneous signal coverage.} Only 3.4--49.5\% of any recording contained usable signal by our detection criteria; the majority of every session was silence, handling noise, or sub-threshold audio, reflecting the practical difficulty of sustained, correctly-positioned, unassisted self-auscultation without any pressure or placement feedback.
\item \textbf{No murmur or arrhythmia screening claim.} While the broader smartphone-PCG literature supports abnormal-sound classification from chest recordings \cite{digitalstethoscope2017,deepauscult2023}, our recordings' SNR and compound sound field make any such claim unsupported here; we make none.
\item \textbf{Self-reported condition labels.} ``Rest'', ``morning'', and ``exercise'' were subject-assigned with no objective verification (e.g., no simultaneous actigraphy or maternal pulse reference), and exercise intensity/timing-since-exertion were not standardized or recorded.
\item \textbf{Small-sample HRV instability.} Our variability metrics (Section~\ref{sec:hrvresults}) use a median of only 6 beats per window; short-sample SDNN/RMSSD estimates are high-variance and should not be compared to clinical short-term-variability thresholds, which assume continuous, multi-minute beat series (Section~\ref{sec:hrvdiscuss}).
\item \textbf{Breathing-rate probe is inconclusive.} The candidate low-frequency rhythm described in Section~\ref{sec:breathingresults} did not significantly separate rest from exercise and may not reflect respiration at all (Section~\ref{sec:breathingfail}); we report it as a negative result and explicitly do not use it anywhere else in this paper's analysis or conclusions.
\end{enumerate}

\section{Future Work}
\label{sec:future}
The most direct next step is a properly instrumented validation study: simultaneous smartphone-microphone abdominal PCG alongside a clinical CTG or Doppler reference would let the maternal/fetal attribution problem be solved with supervision, and let the dual-band heuristic (Section~\ref{sec:dualband}) be replaced or validated against a learned or model-based blind source separation method (e.g., ICA, empirical mode decomposition, or a multi-lag covariance/eigenvalue approach analogous to \cite{comparative_fpcg_ctg_2017}, adapted to the single-channel case). Continuous, multi-minute recordings -- rather than the short clips analyzed here -- are also a prerequisite for any clinically meaningful STV/LTV estimate (Section~\ref{sec:hrvdiscuss}). A multi-microphone accessory, or two phones at different abdominal positions, would reintroduce the spatial diversity purpose-built multi-channel fPCG systems rely on \cite{comparative_fpcg_ctg_2017,fourchannel_validation_2018}. A guided-recording interface -- quality feedback, a contact-pressure indicator, breath-hold prompts -- would likely improve the active-fraction statistics of Section~\ref{sec:results}, the single largest practical bottleneck here. The breathing-rate negative result (Section~\ref{sec:breathingresults}) suggests the smartphone's built-in accelerometer, fused with the audio channel, may be a more direct respiration sensor than acoustic envelope modulation alone. Larger, more diverse cohorts are also needed, and -- in line with the motivation in Section~\ref{sec:motivation} -- future validation work would be especially valuable if conducted in the resource-limited settings this approach ultimately aims to serve, rather than exclusively in well-resourced clinical trial sites.

\section{Conclusion}
\label{sec:conclusion}
Motivated by the stark, well-documented gap between where maternal and fetal monitoring is most needed and where it is currently available \cite{un_sdg3_2026}, and by the near-universal reach of the smartphone in the very regions carrying that burden \cite{gsma_findex_2025}, we investigated whether a consumer smartphone's own microphone, placed passively on the lower maternal abdomen with no additional hardware or emitted signal, can support longitudinal, self-collected tracking of maternal and fetal cardiac rhythm in late pregnancy. Using 28 unsupervised recordings from subject Z over the final five weeks of an uncomplicated pregnancy and a physiologically-informed dual-band decomposition, we recovered a fast, regular, fetal-plausible rhythm ($160.3\pm11.5$~bpm) in 25/28 recordings, stable across maternal rest, morning, and exercise, alongside a slower, noisier maternal-plausible rhythm ($82.5\pm17.9$~bpm) in 14/28. Beat-to-beat variability extracted from both bands, and a further probe for a third, breathing-rate band, both proved on inspection to be dominated by small-sample noise or to show no significant rest/exercise separation -- limitations we surface rather than paper over. These findings are consistent with normal maternal-fetal cardiac status throughout the observed window and the pregnancy's uncomplicated outcome, but rest on an unvalidated frequency-based heuristic and a single subject. We report the pipeline and dataset to responsibly bound the claims of a future validated study, and position passive, commodity-hardware smartphone PCG -- precisely because it requires nothing beyond a device most people on Earth already own -- as an under-explored complement to the active-Doppler apps that currently dominate this space, and as one small, deliberately modest step toward AI-assisted health tools for the communities least served by existing maternal-fetal monitoring technology.

\section*{Acknowledgment}
The author thanks the study subject Z for the sustained, unsupervised, multi-week data collection effort that made this analysis possible.

\footnotesize{
\bibliographystyle{IEEEtran}
\bibliography{refs}}

\end{document}